\documentclass[aps,prb,twocolumn,showpacs,superscriptaddress]{revtex4-1} 
\usepackage{graphicx} 
\usepackage[caption=false]{subfig}
\usepackage{dcolumn}   
\usepackage{bm}  
\usepackage{epstopdf}       
\usepackage{amssymb}
\usepackage{amsmath}   

\begin{document}

\title{The interplay between the structural and magnetic properties of vanadium dioxide from first principles}
\author{J.~M.~Booth}
\email{jamie.booth@rmit.edu.au}
\affiliation{CSIRO Manufacturing and Materials, Clayton VIC 3168, Australia}
\affiliation{Chemical and Quantum Physics, School of Applied Sciences, RMIT University, Melbourne VIC 3001, Australia}

\author{D.~W.~Drumm}
\affiliation{Chemical and Quantum Physics, School of Applied Sciences, RMIT University, Melbourne VIC 3001, Australia}

\author{P.~S.~Casey}
\affiliation{CSIRO Manufacturing and Materials, Clayton VIC 3168, Australia}

\author{J.~S.~Smith}
\affiliation{Chemical and Quantum Physics, School of Applied Sciences, RMIT University, Melbourne VIC 3001, Australia}

\author{S.~P.~Russo}
\affiliation{Chemical and Quantum Physics, School of Applied Sciences, RMIT University, Melbourne VIC 3001, Australia}

\date{\today}

\begin{abstract}
The effects of the spin and lattice degrees of freedom on the electronic structure of M1 vanadium dioxide are explored using a quasiparticle description. Contraction of the inter-vanadium spacing of the Peierls pairings stabilizes bonding electrons, reducing polarizability and thus widening the band gap. Increases of this inter-vanadium spacing of as little as 1 \% reduce this stabilization, resulting in a crossover to ferromagnetic behaviour accomplished by half of the valence electrons inhabiting the leading edge of the conduction band in localized atomic-like orbitals, as the antiferromagnetic order becomes unstable with respect to rearrangement according to Hund's first rule. The data indicates that the magnetic structure of M1 vanadium dioxide may be finely balanced; the antiferromagnetic order is a consequence of the overlapping nuclear potential of the Peierls pairs, and input which disrupts this will have a significant effect on magnetic properties.
\end{abstract}

\maketitle

\section{Introduction}
The potential of phenomena arising from strongly correlated materials, such as Colossal Magnetoresistance, Metal-Insulator Transitions and High Temperature Superconductivity, for the production of a new generation of devices can hardly be overstated. \cite{Dagotto2005} The possibilities of high temperature superconductivity in particular have created an explosion of interest in the properties of strongly correlated oxides. \cite{Basov2011} One such correlated oxide, M1 vanadium dioxide, has long been the subject of intense practical and theoretical interest since the discovery of its metal-insulator transition (MIT)by Morin. \cite{Morin1959,Goodenough1971,Mott1974,Pouget1975,Zylberstein1975,Imada1998} 
\begin{figure}[h!]
   \centering
    \subfloat{{(a)}\includegraphics[width=0.45\columnwidth]{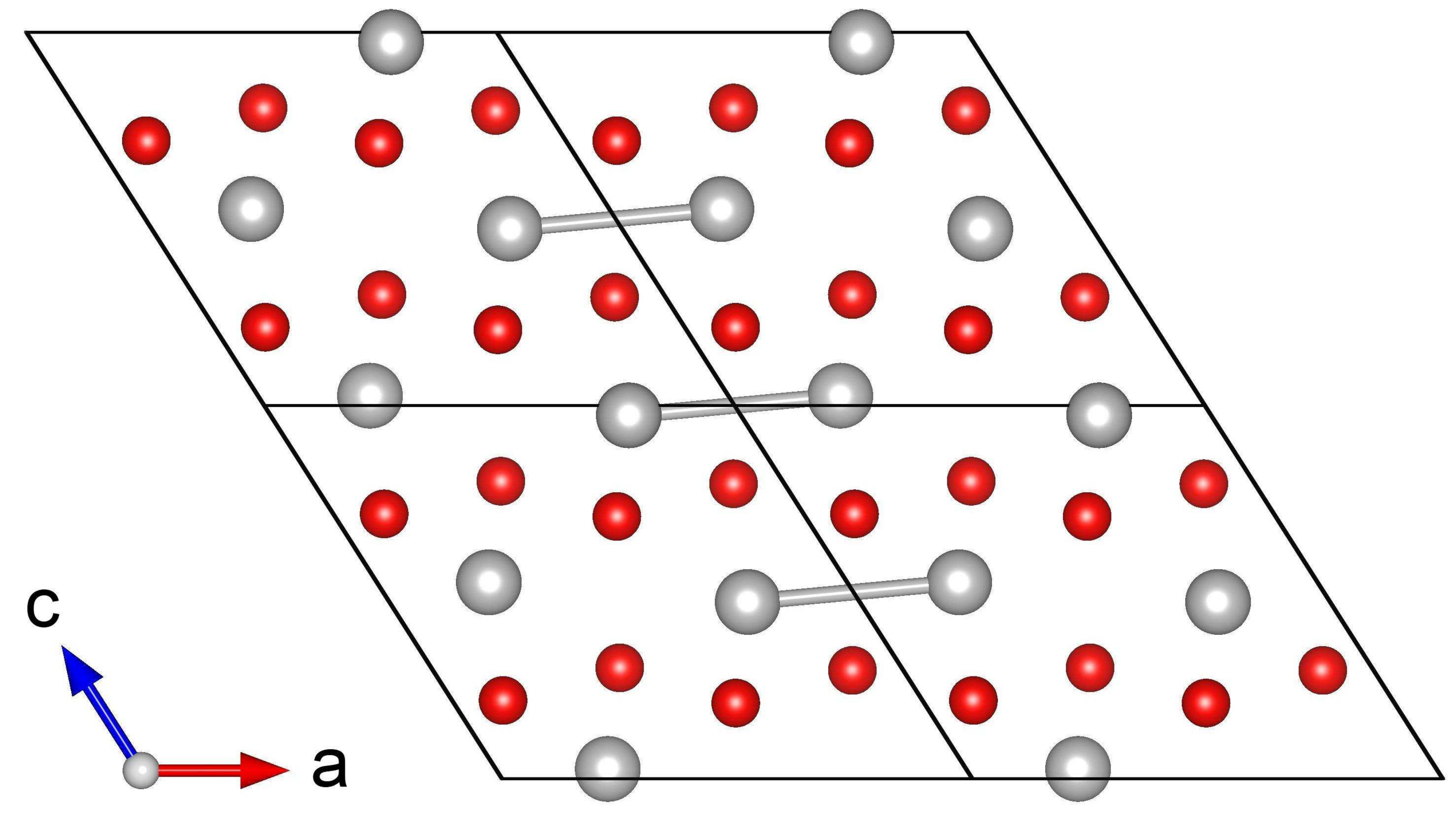}}   \subfloat{{(b)}\includegraphics[width=0.35\columnwidth]{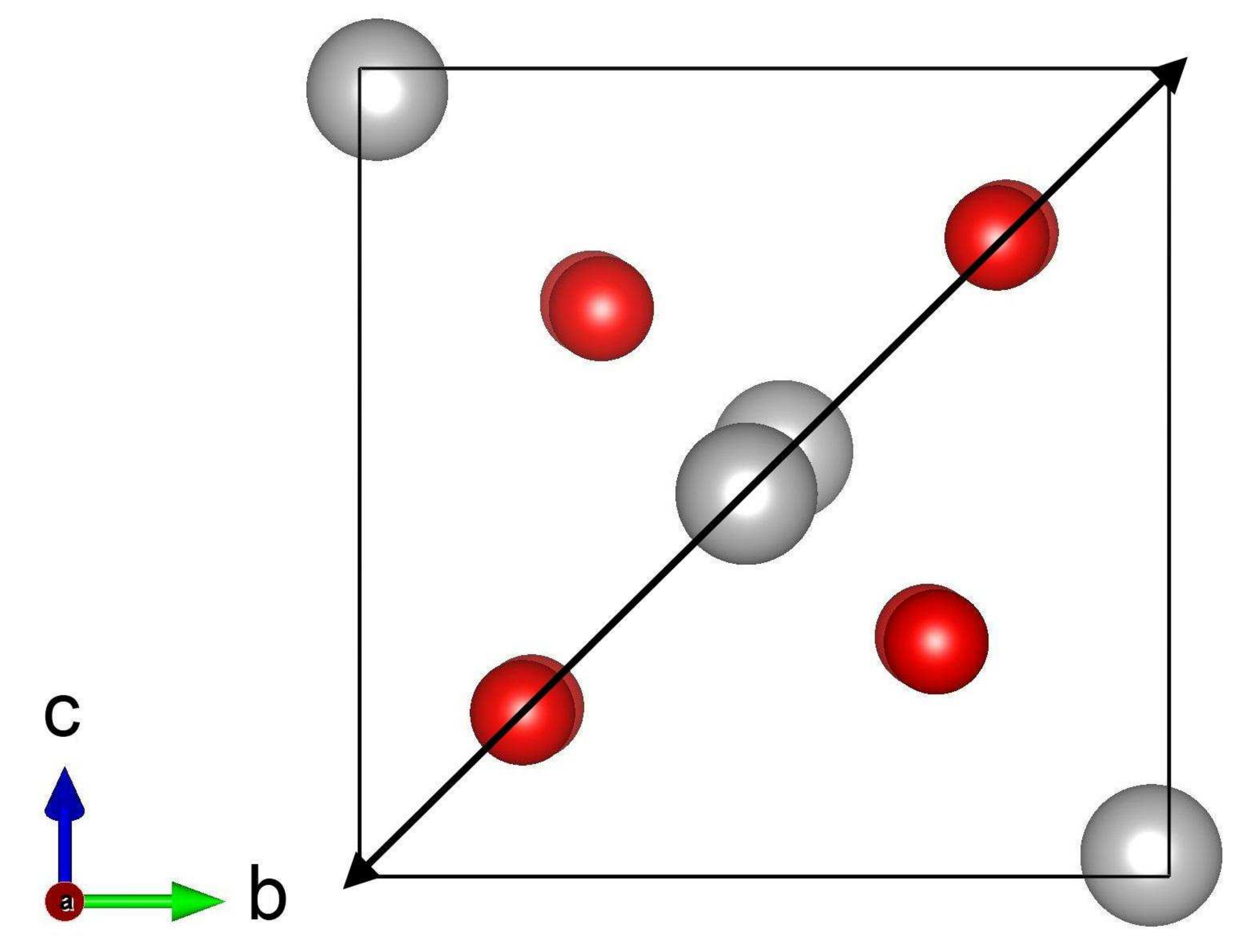}}
    \caption{Structure of M1 vanadium dioxide viewed down a) the b-axis, the bonds indicate the Peierls pairings and b) the a-axis, the diagonal arrow indicates the orientation of the $(0\bar{1}1)$ plane which the charge densities of Figure 4 are taken from. Vanadium atoms are grey and oxygen atoms are red.}
\end{figure}

This transition from the high temperature metallic tetragonal structure to the low temperature insulating monoclinic structure is accomplished via Peierls pairing of the vanadium atoms combined with an antiferroelectric twist. \cite{Goodenough1971} Modern computational techniques such as Dynamical Mean Field Theory \cite{Kotliar2006} and the GW approximation \cite{Hedin1965} have shed considerable light on the nature of the insulating phase, \cite{Tomczak2008,Gatti2007,Belozerov2012} indicating that the band gap results from Peierls localization \cite{Peierls1955} of the $d$-band electrons driven by strong correlations in the metallic structure.

In recent years, nanofabrication approaches have been developed which can generate structures suitable for high performance devices, and authors have begun to exploit the MIT of VO$_2$ nanobeams and films for sensing applications, \cite{Wu2006,Zhou2008,Sohn2009,Guo2011} while Nakano \textit{et al.} demonstrated a new type of Field Effect Transistor. \cite{Nakano2012} What these approaches have in common, is that modulation of the characteristics of the MIT is achieved by inputting stress or strain. 

Despite this interest however, the effect of changes in lattice structure on the electronic structure in the context of strong correlations has not been established. In particular, while insulating VO$_2$ is antiferromagnetic, it is unclear how robust this ordering is with respect to the spacing of the Peierls configuration. In this study, spin-resolved GW calculations are employed to determine this dependency. It is found that the electronic structure can be significantly modified by small changes in the lattice. In particular, strain input which increases the Peierls spacing may result in a transition from antiferromagnetic to ferromagnetic behaviour.

\section{Methods}
\subsection{Structures and Geometry Relaxations}
The lattice parameters of the M1 structure used were those obtained by Andersson. \cite{Andersson1954} All calculations were performed using the Vienna Ab Initio Simulation Package. (VASP)\cite{Kresse1996} Geometry relaxations were performed using Density Functional Theory (DFT) in the Generalized Gradient Approximation (GGA), using PBE06 functionals \cite{Perdew1996} and Methfessel and Paxton smearing. \cite{Methfessel1989} Single point DFT calculations employed the tetrahedron method with Bloechl corrections. \cite{Bloechl1994} The intermediate structures (labelled $\frac{1}{4}$, $\frac{1}{2}$ and $\frac{3}{4}$) were generated by calculating vectors describing the change in both lattice parameters and atomic positions resulting from the geometry relaxation, multiplying them by the labels listed, and adding them to the geometry relaxed structure. Thus, the structures consist not only of intermediate lattice constants, but also atomic positions. The structures generated for Figure 4 are simple substitutions of either lattice constants, or lattice constants and atomic positions, and are explained in the discussion. Likewise, increases in the Peierls pairing distances were generated by changing the fractional coordinates of the vanadium atoms, holding all other parameters constant.
\subsection{Quasiparticle Calculations}
Quasiparticle shifts were calculated using the G$_0$W$_0$ implementation of Shishkin and Kresse. \cite{Shishkin2006} Initially density functional theory \cite{Kohn1965} (DFT) with the Generalized Gradient Approximation to exchange and correlation \cite{Perdew1996} was used to converge the wavefunctions on an 8-atom cell, using a 6$\times$6$\times$6 Monkhorst Pack \cite{Monkhorst1976} k-point grid, the Brillouin Zone integration method of Bloechl \textit{et al.} \cite{Bloechl1994} and an exact diagonalization of the Hamiltonian. The GW approximation was used to calculate the eigenvalue shifts due to the electron self energies on a 25 point frequency grid, which was found to give results that were for all intents and purposes, identical to grids of 30 - 50 frequencies, while significantly reducing computation time. 

Neglecting vertex terms the first order self energy expansion is written:\cite{Hedin1965}
\begin{equation}
\Sigma=iGW
\end{equation}
The screened interaction in momentum-space is given by:\cite{Guiliani2005}
\begin{equation}
W(\mathbf{q},\omega)=\frac{\nu_\mathbf{q}}{1-\nu_\mathbf{q}\chi(\mathbf{q},\omega)}
\end{equation}
The $\chi(\mathbf{q},\omega)$ term is the independent particle polarizability (independent as we are using the Random Phase Approximation) which can be written:\cite{Hybertsen1987, Shishkin2006}
\begin{multline}
	\chi^{0}_\mathbf{q}(\mathbf{G,G^{\prime}},\omega)=\frac{1}{\Omega}\Sigma_{nn^{\prime}\mathbf{k}}2w_{\mathbf{k}}(f_{n^{'}\mathbf{k-q}}-f_{n\mathbf{k})}\\\times\frac{\langle\psi_{n^{'}\mathbf{k-q}}|e^{-i\mathbf{(q+G)r}}|\psi_{n\mathbf{k}}\rangle\langle\psi_{n\mathbf{k}}|e^{i\mathbf{(q+G^\prime)r^\prime}}|\psi_{n^{'}\mathbf{k-q}}\rangle}{\omega+\epsilon_{n^{\prime}\mathbf{k-q}}-\epsilon_{n\mathbf{k}}+i\eta\text{sgn}[\epsilon_{n^{\prime}\mathbf{k-q}}-\epsilon_{n\mathbf{k}}]}
\end{multline}
This polarizability modifies the Coulomb potential, $\nu(\mathbf{q})$, to infinite order, which is summed to give the first order expansion in the \textit{screened interaction} in equation (2).\cite{Guiliani2005} Equation (3) depends on three main factors: it is directly proportional to difference of the occupancies of the bands on the k-point mesh, $f_{n\mathbf{k}}$, and the exchange charge density (the angled bracket product term), while it is inversely proportional to the difference in the energies of the two states separated by $\textbf{q}$, $\epsilon_{n^{\prime}\mathbf{k-q}}$-$\epsilon_{n\mathbf{k}}$ where $n$ and $n^{\prime}$ label the bands. 

This self-energy operator can be used in place of the usual DFT exchange-correlation potential to generate the band eigenvalues using the quasiparticle equation (QP):\cite{Shishkin2006}
\begin{equation}
E^{QP}_{n\mathbf{k}}=Re[\langle\psi_{n\mathbf{k}}|T+V_{n-e}+V_H+\Sigma(E^{QP}_{n\mathbf{k}})|\psi_{n\mathbf{k}}\rangle]
\end{equation}

A cut-off energy for the response function calculation of 200 eV was used for all calculations.  

\section{Results and discussion}
Figure 2 compares the spin-resolved electronic densities of states (eDOS) obtained from both density functional theory and the GW approximation. Figures 2a-b are the DFT results for the experimental structure parameters, and the geometry relaxed structure respectively. Figures 2c-d are the GW results for the experimental and relaxed structures respectively. As expected, DFT predicts the experimental structure to be metallic, rather than the observed semiconductor, which has been confirmed by numerous studies. \cite{Wentzcovich1994,Eyert2002} In addition, the structure is predicted to be Ferromagnetic, again in contrast to experiment. Of note is the splitting of the high energy $d$-band states near the Fermi level (E$_F$), which we return to later. Conversely, the relaxed structure is predicted by DFT to be antiferromagnetic, and semiconducting; the calculated eDOS exhibits a clear splitting of the $d$-band states at E$_F$, although the gap width is significantly underestimated with respect to the experimentally determined value of 0.65 eV,\cite{Eyert2002} manifesting simply as zero density of states at E$_F$. Table 1 lists the parameters of the DFT relaxed structure (labeled ``GR") and the experimental structure (labelled ``EXP") and the percentage differences between them ($\Delta$). 
\begin{figure*}[t!]
  \centering
  \includegraphics[width=1.5\columnwidth]{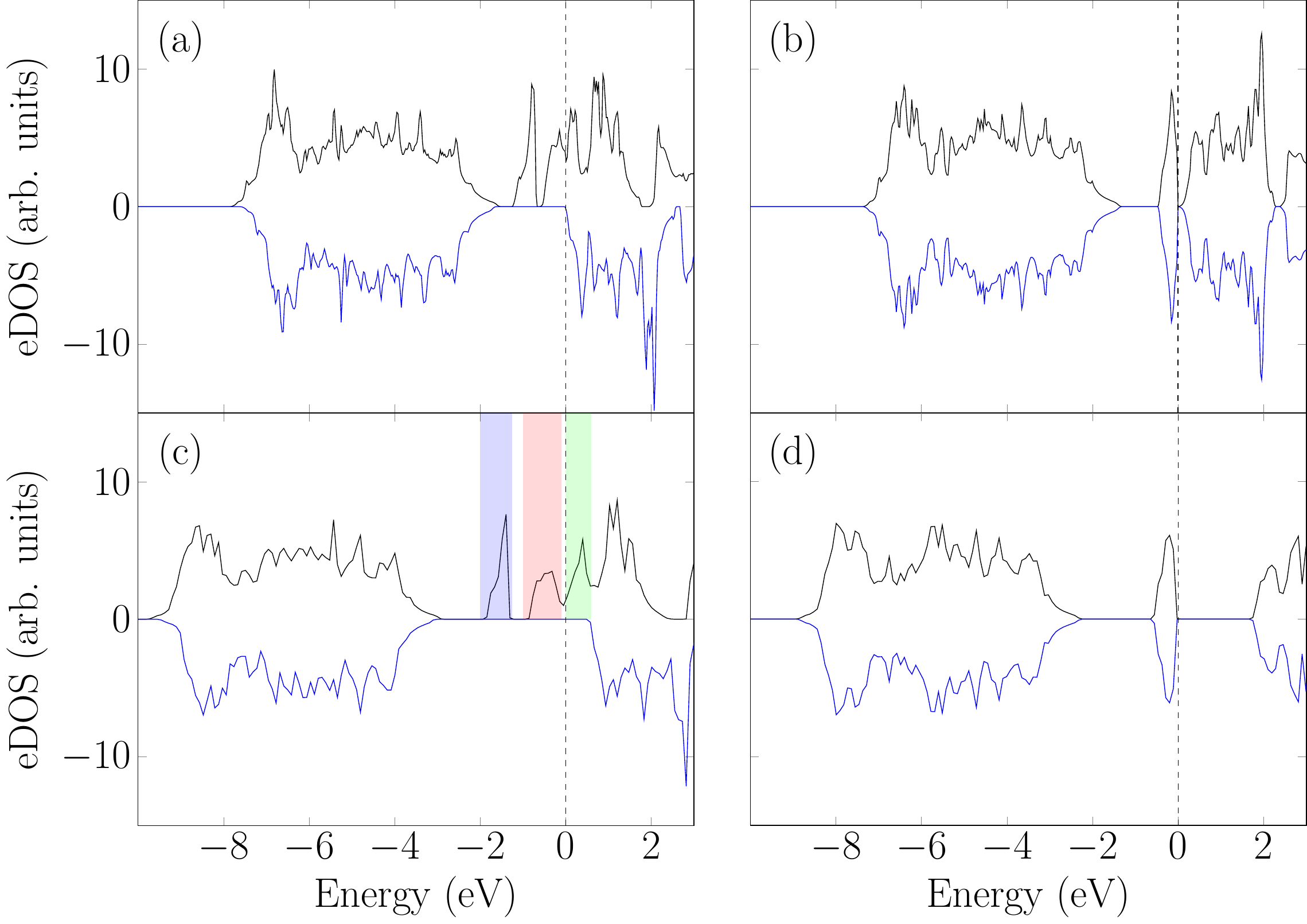}
  \caption{\label{fig:eDOSs} a) DFT eDOS of the experimental (EXP) parameters from Andersen, \cite{Andersson1954} b) DFT eDOS of the structure obtained from a DFT geometry relaxation, c) GW eDOS of the EXP structure and d) GW eDOS of the geometry relaxed structure all relative to the Fermi level (dashed line).}
\end{figure*}
\begin{table}[ht!]
  \centering
  \begin{tabular}{ccccccc}
    \hline
    \hline
     & GR & $\frac{1}{4}$ & $\frac{1}{2}$ & $\frac{3}{4}$ & EXP & $\Delta$ \%\\
    \hline
    $\mathbf{a}$ ($\mathrm{\AA}$)		& 5.656 & 5.680 & 5.704 & 5.728 & 5.752 & -1.67 \\
    $\mathbf{b}$ ($\mathrm{\AA}$) 	& 4.601 & 4.585 & 4.569 & 4.554 & 4.538 &	1.40 \\
    $\mathbf{c}$ ($\mathrm{\AA}$)		& 5.410 & 5.403 & 5.396 & 5.389 & 5.383 & 0.50\\
    $\beta$ ($^\circ$)		& 122.07 & 122.21 & 122.36 & 122.50 & 122.65 & -0.47 \\
    V-V (S) ($\mathrm{\AA}$)		& 2.520 & 2.545 & 2.569 & 2.594 & 2.619 & -3.77 \\
    V-V (L) ($\mathrm{\AA}$)	& 3.174 & 3.172 & 3.170 & 3.168 & 3.166 & 0.27 \\
    Gap (eV) & 1.66 & 0.80 & 0.65 & metal & metal & N/A \\
    \hline
  \end{tabular}
      \caption{Unit cell parameters and inter-vanadium spacings along the a-axis of the five structures used in this study and the \% difference between the EXP and GR structures, $\Delta$.}
\end{table}

The data indicates that the DFT relaxation results in a contraction along the a-axis, which combines with an almost 4 \% decrease in the spacing between the Peierls paired atoms (V-V (S)), partially compensated for by a 0.27 \% increase in the long V-V distance (V-V (L)). The other crystallographic directions however experience slight expansions, while the $\beta$ angle contracts slightly. Overall, the changes are minor, and can be summarized as a slight exaggeration of the Peierls distortion. Thus, the data of Figures 2a-b and Table 1 suggest that despite its reputation for failure with respect to strongly correlated materials, DFT can achieve reasonable agreement with experiment with respect to the structural parameters of undoped M1 VO$_2$ structures. The electronic structure however is in rather poor agreement. While the prediction of antiferromagnetic behaviour is correct, the band gap is significantly underestimated, which makes the determination of the effects of structural distortions on the electronic structure extremely uncertain.

The frequency-dependent GW data of Figures 2c-d also paint an interesting picture. The EXP structure is again calculated to be ferromagnetic, with the $d$-band states split into two spin up peaks, separated by a gap of $\sim$0.35 eV, while the high energy peak appears to form a doublet structure which straddles E$_F$. The geometry relaxed structure, in almost total contrast, is antiferromagnetic. However the GW data exhibits a band gap of 1.66 eV, which is more than double the experimentally determined value. Table 1 suggests that this significant discrepancy in electronic structure is essentially the result of an $\sim$3.8 \% reduction in the internuclear Peierls spacing. The data therefore indicates that the magnetic and insulating behavior is extremely structurally dependent, and presents an opportunity to examine the interplay between the structural and spin degrees of freedom. Figure 2 also suggests that a structure with electronic behavior which agrees with experiment lies between the EXP and GR structures.
\begin{figure}[ht!]
\centering
\includegraphics[width=0.7\columnwidth]{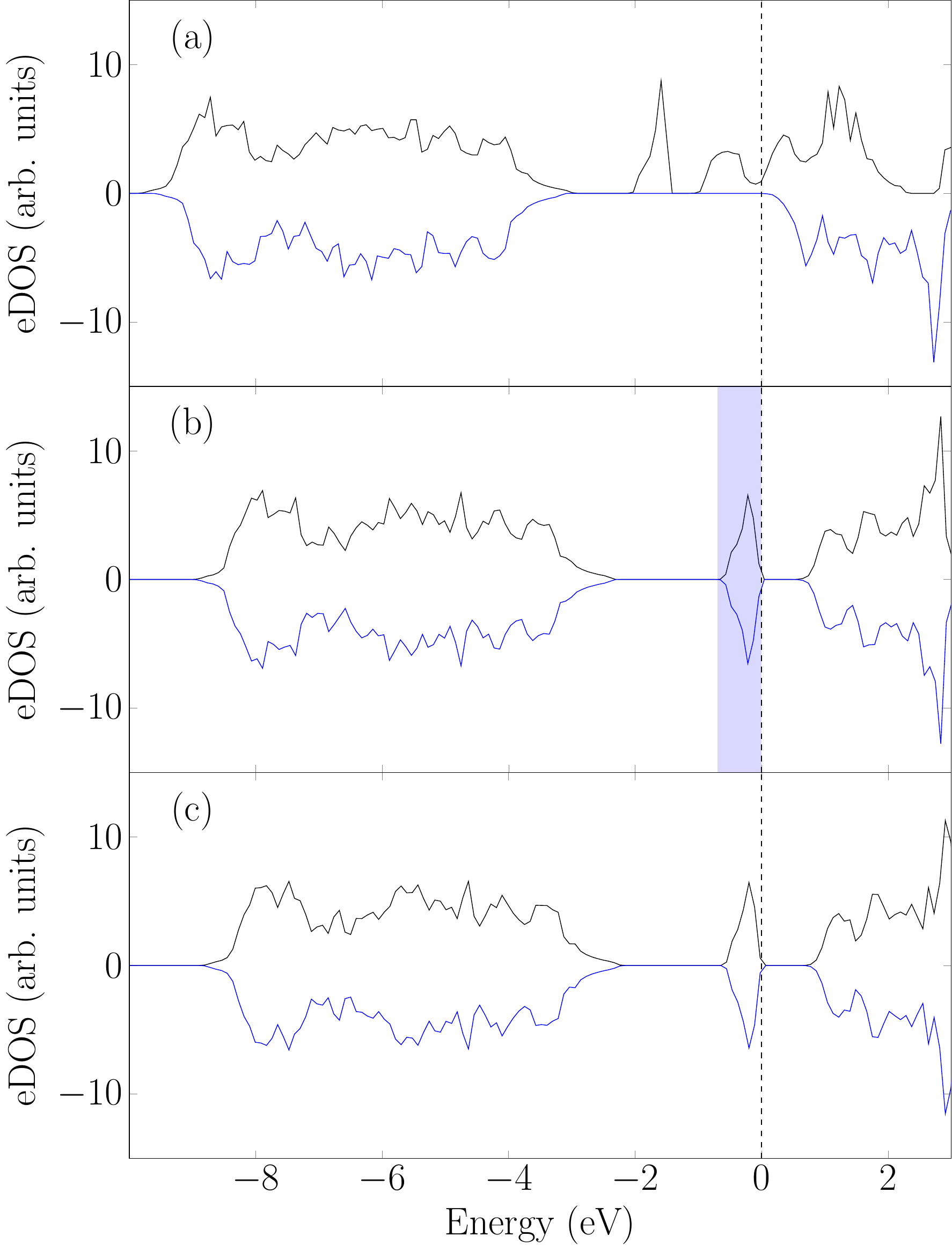}
\caption{GW eDOS of the a)  $\frac{3}{4}$ structure, b) $\frac{1}{2}$ and c) $\frac{1}{4}$ structure all relative to the Fermi level (dashed line).}
\end{figure}

\begin{figure*}[t!]
  \centering
  \includegraphics[width=1.5\columnwidth]{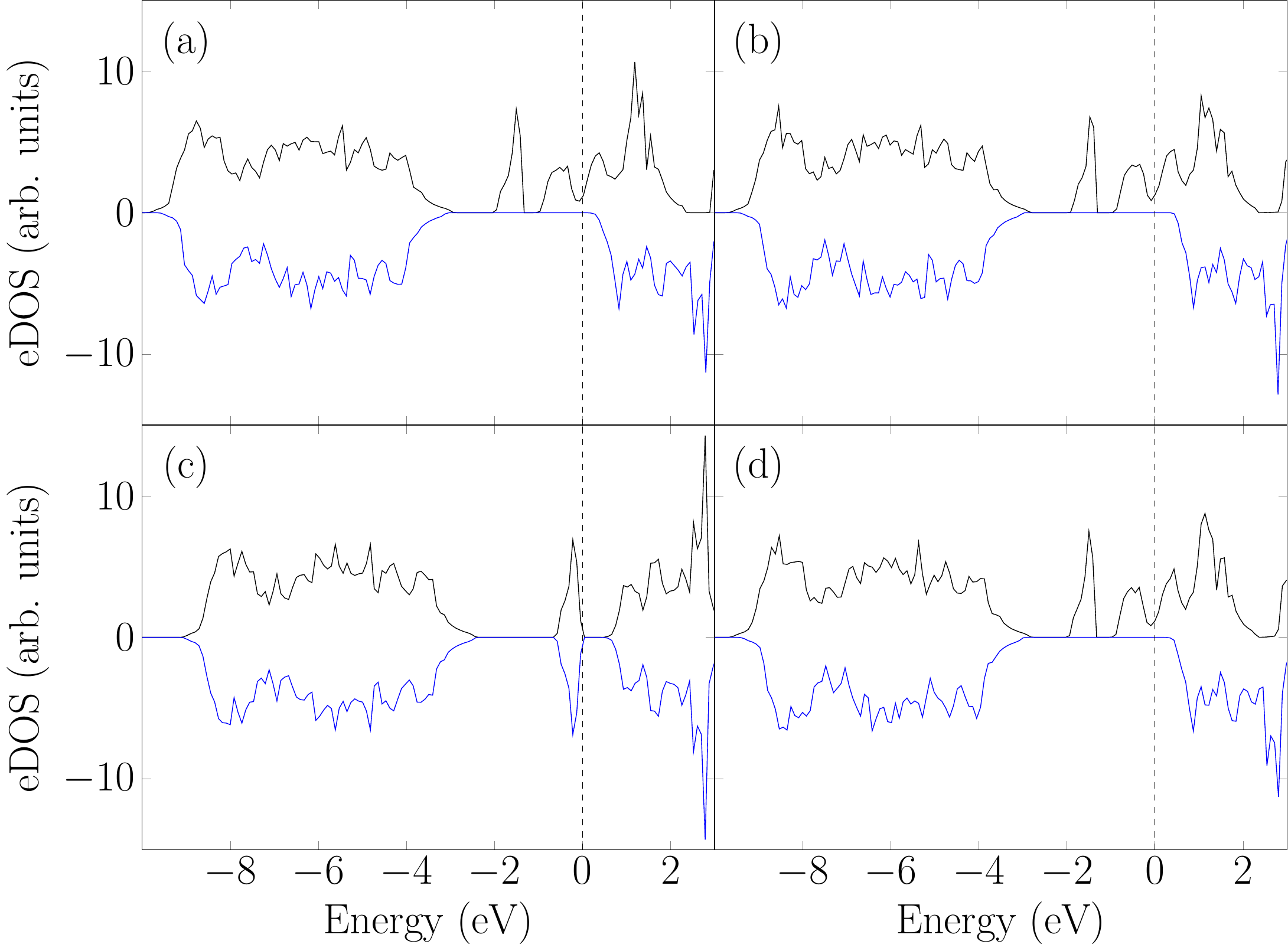}
  \caption{\label{fig:eDOSs} GW densities of states of a) a structure identical to the experimental structure, however with the a-axis length equal to the $\frac{1}{2}$ structure, b) this structure retains the experimental a-axis length, but has the same b- and c-axis lengths as the $\frac{1}{2}$ structure, c) this structure has both the a-axis length of the $\frac{1}{2}$ structure and the a-axis coordinates for the atomic positions (i.e. it retains the b- and c-axis lengths and coordinates of the experimental structure) d) this form has both the b- and c-axis lengths and atomic positions of the $\frac{1}{2}$ structure (i.e. it the a-axis length and atomic positions are the same as those of the experimental structure).}
\end{figure*} 

In order to investigate this, three intermediate structures were obtained by linearly interpolating between the EXP and GR structures, which (setting GR as zero and EXP as one) were named $\frac{1}{4}$, $\frac{1}{2}$ and $\frac{3}{4}$. The unit cell parameters of these structures are also listed in Table 1, along with the corresponding band gaps. The $\frac{3}{4}$ form exhibits a very similar electronic structure to the EXP form, being ferromagnetic and exhibiting splitting of the $d$-band states below the Fermi level into two peaks. The $\frac{1}{2}$ and $\frac{1}{4}$ structures on the other hand are antiferromagnetic insulators, which exhibit band gaps of approximately 0.65 eV and 0.80 eV respectively. Thus the $\frac{1}{2}$ structure exhibits a band gap which agrees most closely with the experimentally determined value of $\sim$ 0.65 - 0.7 eV. \cite{Eyert2002} This structure differs from the ``EXP" structure by half the percentages listed in column 7 of Table 1. Therefore, most of the changes represent expansions or contractions of less than 1 \% apart from the exaggeration of the Peierls pairing. However, even this represents only a $\sim$1.9 \% decrease in the short V-V distance, indicating close agreement between theory and experiment.

The combinations of Figures 2d, 3a-c and the data of Table 1 therefore reveal the dependency of the band gap on structural parameters, and in particular the Peierls pairing distance. As the Peierls pairing distance decreases (e.g. going from the $\frac{1}{2}$ to the $\frac{1}{4}$ structure), equivalent to a compressive stress, the magnitude of the band gap increases. This echoes resistivity data which revealed an increase in the resistance of the insulating phase under compression along the monoclinic a-axis.\cite{Muraoka2002} This decrease in spacing will increase the nuclear potential overlap between the vanadium atoms, and therefore the potential energy components corresponding to the electron-nuclear interaction will become more negative, lowering the potential energy. This in turn will affect the polarizability of Equation (3) by increasing the energy separation between these states and the empty states given by the denominator, reducing polarizability and increasing the magnitude of the band gap. Going the other way, from the $\frac{1}{2}$ to the $\frac{3}{4}$ structure, the gap closes completely. The increase in the Peierls spacing, corresponding to a tensile strain, results in a lower stabilization of the repulsion of the Peierls paired electrons, and consequently, as the splitting of the peaks in the eDOS indicates, one of the Peierls bonding electrons is released into the conduction band. Therefore, under strain we expect a significant decrease in resistivity of the insulating phase, once again confirmed by experiment.\cite{Gregg1997,Muraoka2002}

With the experimental dependency of the band gap now reproduced, we are in a position to determine the exact nature of the effects of strain perturbations. Figure 4 displays the eDOSs of four modifications of the experimental structure: a) a structure identical to the experimental, except that the a-axis length is equal to the $\frac{1}{2}$ structure, b) a structure that retains the experimental a-axis length, but has the same b- and c-axis lengths as the $\frac{1}{2}$ structure. Figures 4c) and 4d) are a little more subtle. Figure 4c) has both the a-axis length of the $\frac{1}{2}$ structure and the a-axis fractional coordinates for the atomic positions (i.e. it retains the b- and c-axis lengths and fractional coordinates of the experimental structure). Figure 4d) similarly has both the b- and c-axis lengths and atomic positions of the $\frac{1}{2}$ structure (i.e. it the a-axis length and atomic positions are the same as those of the experimental structure).

\begin{figure}[ht!]
\centering
\includegraphics[width=\columnwidth]{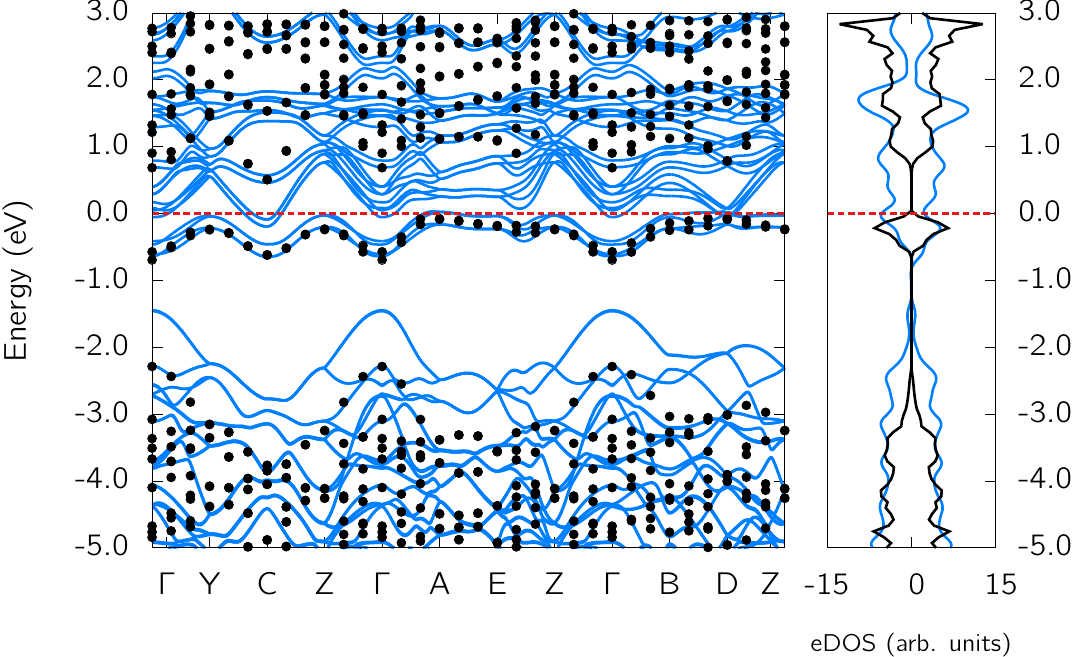}
\caption{DFT and GW Band structures and corresponding densities of states of the $\frac{1}{2}$ structure, which has a Peierls spacing of 2.569 $\mathrm{\AA}$. For the GW data, filled black circles represent up spins, while down spins are empty circles (in this Figure the bands are degenerate so this is difficult to see). The DFT bands are blue for both spin up and down. The GW DOS is plotted in black while the DFT DOS is plotted in blue. }
\end{figure}

\begin{figure}[ht!]
\centering
\includegraphics[width=\columnwidth]{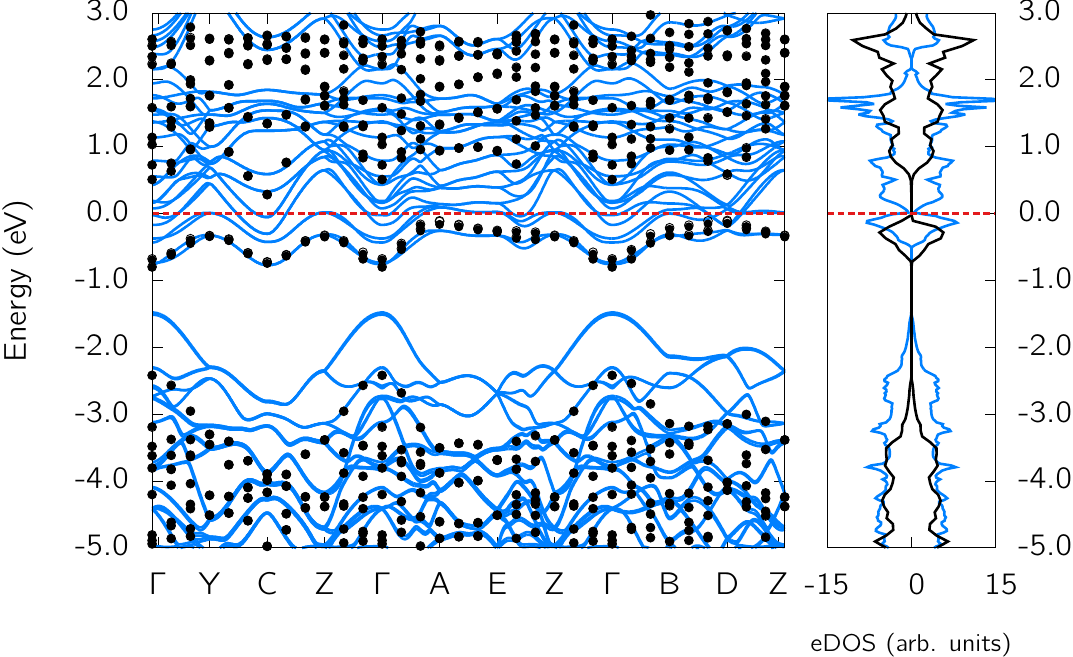}
\caption{Band structure of the $\frac{1}{2}$ structure with a Peierls spacing of 2.587 $\mathrm{\AA}$. Symbol and line conventions are as per Figure 5.}
\end{figure}

\begin{figure}[ht!]
\centering
\includegraphics[width=\columnwidth]{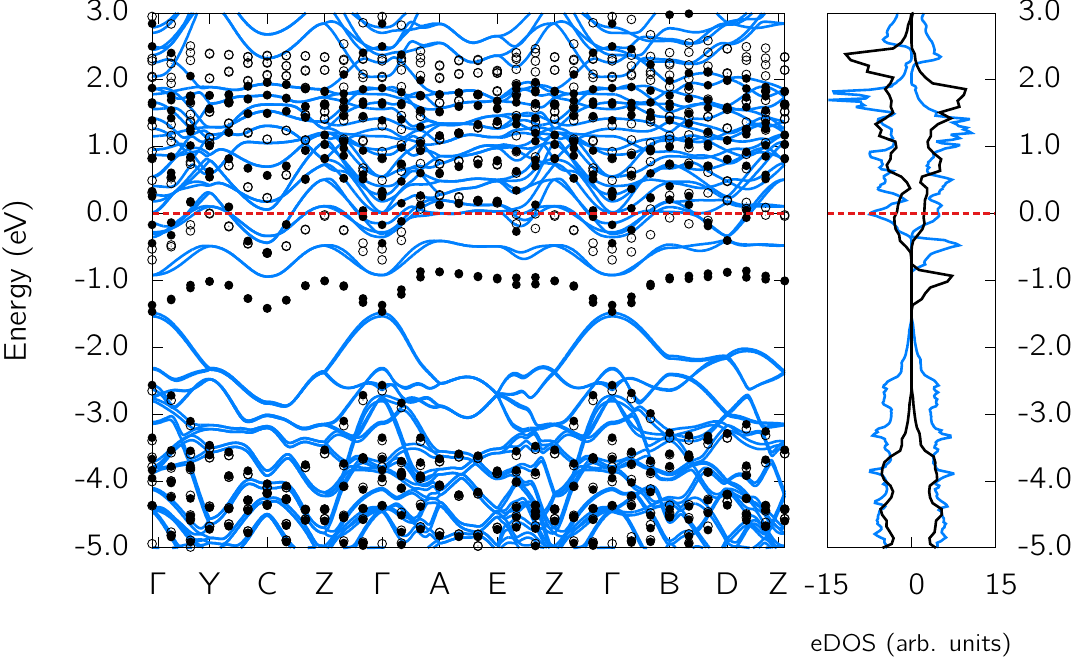}
\caption{Band structure of the $\frac{1}{2}$ structure with a Peierls spacing of 2.597 $\mathrm{\AA}$. Symbol and line conventions are as per Figure 5.}
\end{figure}

\begin{figure}[ht!]
\centering
\includegraphics[width=\columnwidth]{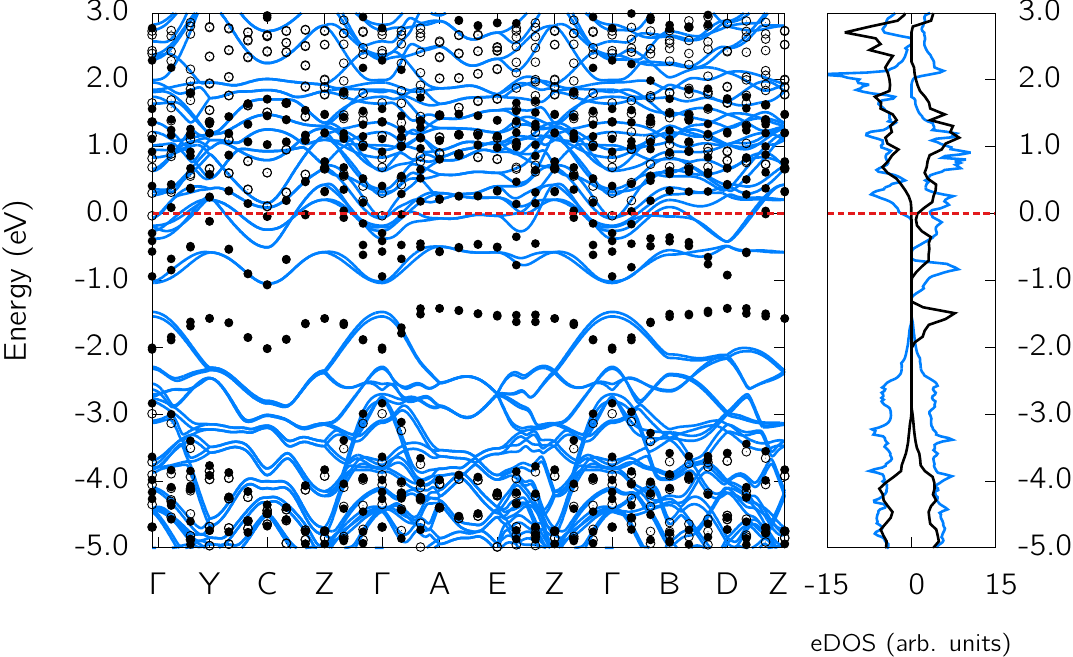}
\caption{Band structure of the $\frac{1}{2}$ structure with a Peierls spacing of 2.606 $\mathrm{\AA}$. Symbol and line conventions are as per Figure 5.}
\end{figure}
Figure 4 reveals that the change from ferromagnetic to antiferromagnetic behavior is solely a consequence of the change in the a-axis length and fractional atomic coordinates. Inputting just the a-axis contraction (Figure 4a), or the b- and c-axis expansion (Figure 4b) has almost no effect on the electronic structure. Similarly, inputting the b- and c-axis expansions together with the associated fractional atomic positions (Figure 4d) does not change the electronic structure. Only Figure 4c shows any change, and thus it is the changes in the short and long inter-vanadium spacing along the monoclinic a-axis which determines the magnetic properties.

Figures 5-8 presents a more detailed investigation of this effect by plotting band structures obtained by gradually increasing the inter-vanadium spacing of the Peierls pairs from the $\frac{1}{2}$ structure. Figure 5 displays the bands of the $\frac{1}{2}$ structure, which as Table 1 indicates has a Peierls spacing of 2.569 $\mathrm{\AA}$. Figures 6-8 correspond to structures with Peierls spacings of 2.587 $\mathrm{\AA}$, 2.597 $\mathrm{\AA}$ and 2.606 $\mathrm{\AA}$ respectively. The band structure of the $\frac{1}{2}$ structure indicates that there is little difference between the band dispersions calculated by DFT and GW. The band splitting is significantly increased when the GW approximation is used however. Most significantly, a band gap opens at the Fermi level however the oxygen $\textit{p}$-bands also shift to considerably lower energy. Also of interest is that the slight splitting of the up and down filled $d$-bands and the low lying conduction bands in the DFT calculation disappears in the GW calculation, with all bands effectively degenerate. This indicates that along with the band gap of the $\frac{1}{2}$ structure more closely matching experiment when using the GW approximation, the spin degeneracy of this structure is also more closely aligned with experiment. Figure 6 indicates that the electronic structure is relatively stable to small changes in the Peierls spacing; the band dispersions are almost identical. The only differences are minor: the band gap has decreased in magnitude from 0.65 eV to 0.5 eV (in agreement with experimental data\cite{Muraoka2002}), and the spin degeneracy has been slightly lifted along the A $\rightarrow$ E and B $\rightarrow$ D high symmetry paths. This results in slight asymmetry in the densities of states of the up and down spins in the $d$-band, however away from E$_F$ the spin degeneracy remains intact.

\begin{figure}[h*t]
    {\centering
    \subfloat{\includegraphics[width=0.45\columnwidth]{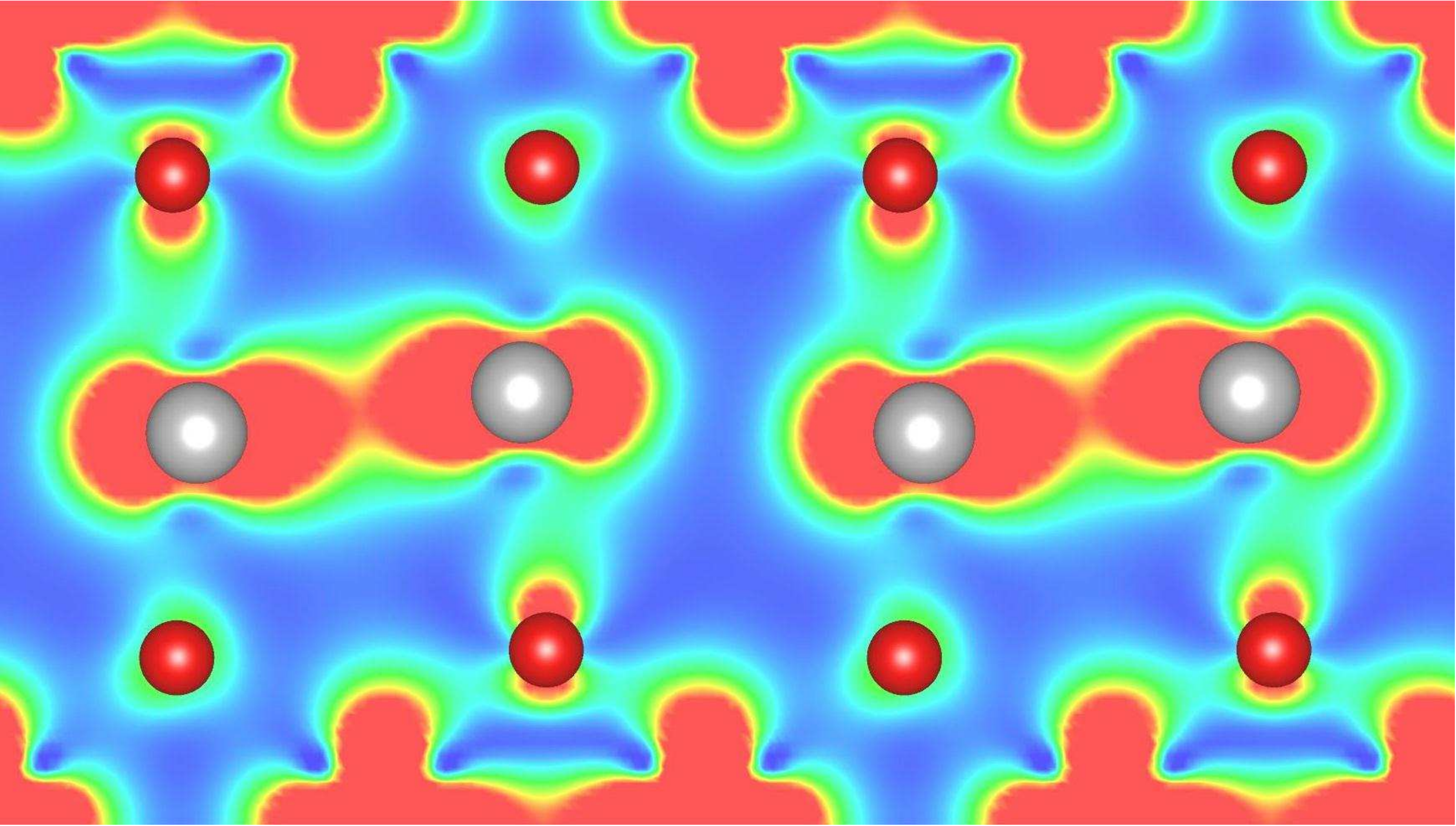}}
    \quad
    \subfloat{\includegraphics[width=0.45\columnwidth]{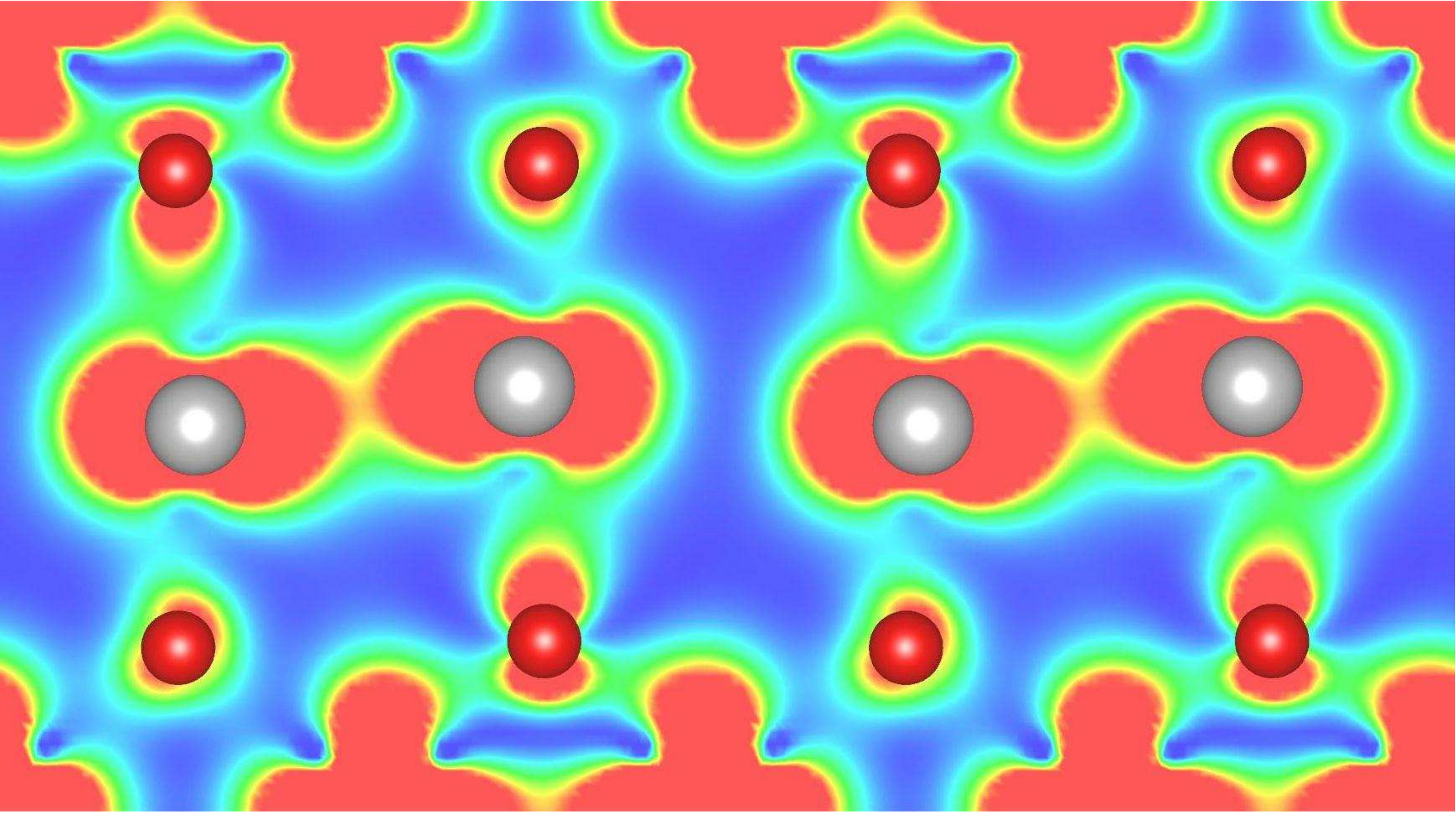}}\\
    \subfloat{\includegraphics[width=0.45\columnwidth]{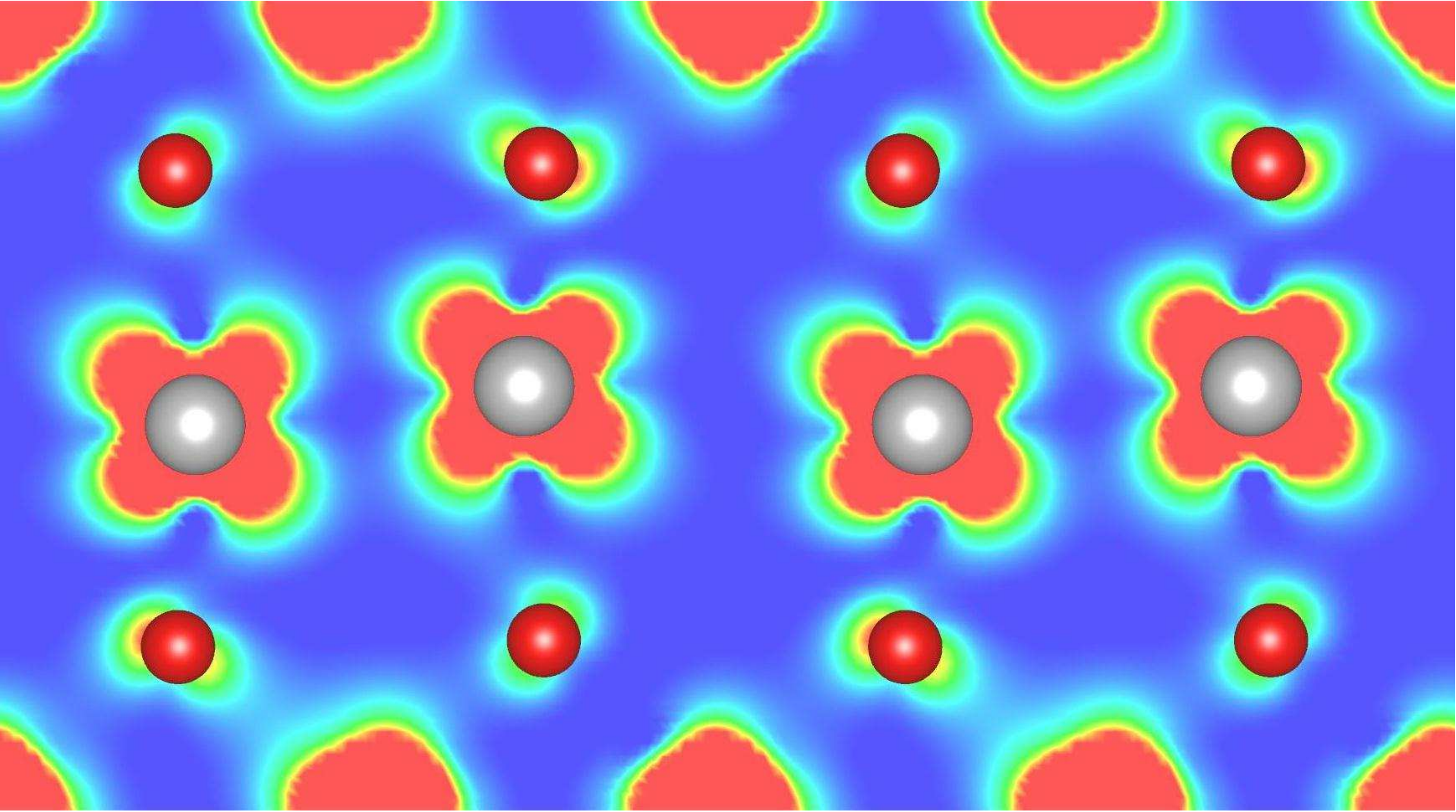}}
    \quad
    \subfloat{\includegraphics[width=0.45\columnwidth]{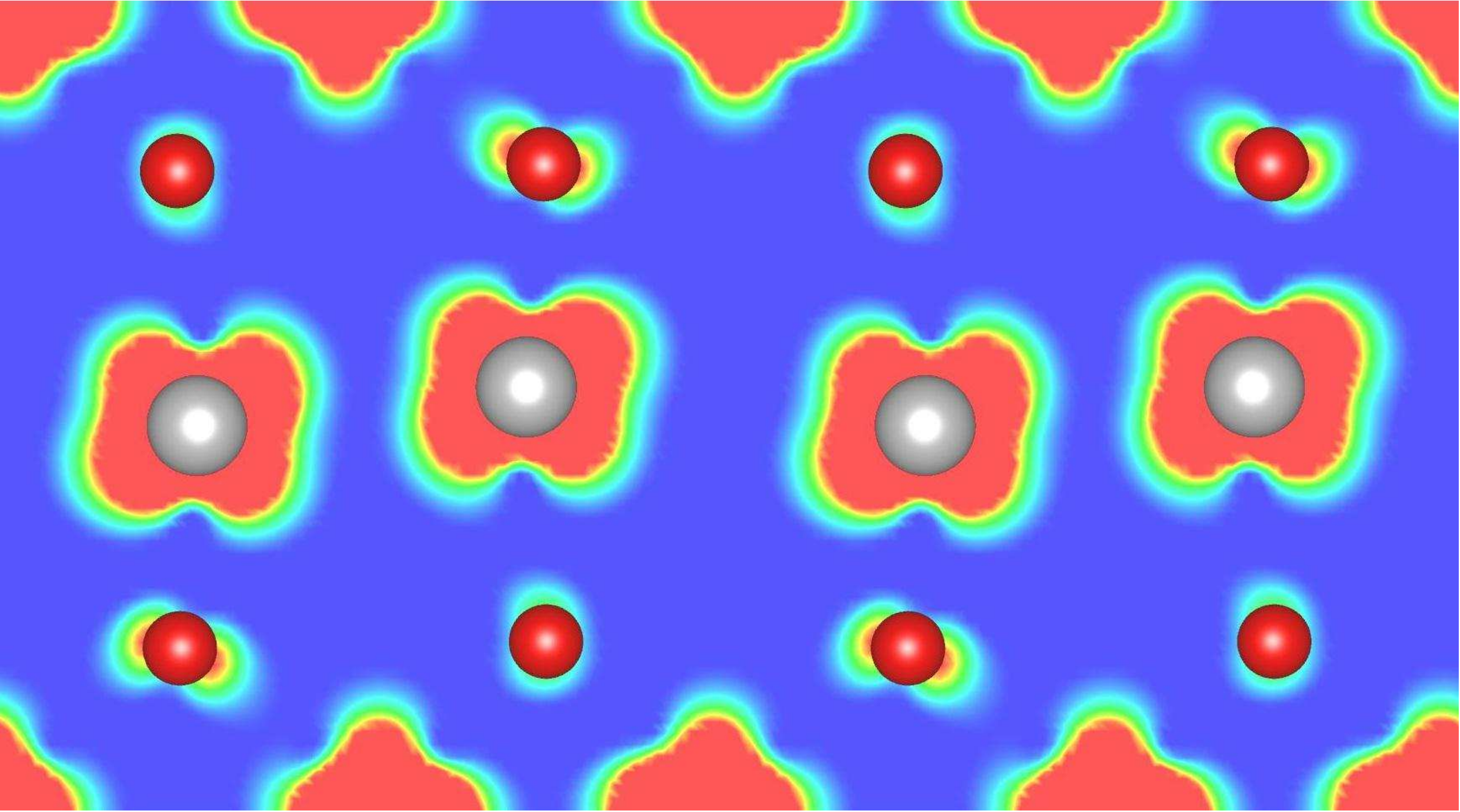}}
    \caption{GW charge densities in the $(0\bar{1}1)$ plane of the a) $d$-band peak of the $\frac{1}{2}$ structure (blue shaded region of Figure 3b, b) the low energy $d$-band peak of the EXP structure (blue shaded region of Figure 2c), c) the high energy $d$-band peak of the EXP structure (red shaded region of Figure 2c) and d) the lowest conduction band of the EXP structure (green shaded region of Figure 2c). Red indicates high density, green indicates intermediate density, while dark blue corresponds to low density. Atomic colors are as per Figure 1.}}
\label{fig:4}
\end{figure}

In Figure 7, it is seen that sufficiently large increases in the Peierls spacing completely lifts the spin degeneracy of the $d$-band. In this structure, the original Peierls bonding band is split in two. The up states shift to lower energy, and while the down states remain in almost the same positions relative to E$_F$, the band width increases slightly due to the most strongly correlated states; the regions of flat dispersions along A $\rightarrow$ E and B $\rightarrow$ D being lifted above the Fermi level. The formation of these hole regions results in conduction band states being filled and dropping below E$_f$ to compensate, dragging the low-lying conduction bands down and closing the gap. Thus we see that as the Peierls spacing increases, the overlapping nuclear potential is no longer strong enough to trap two electrons in a homopolar bond. One of the spin states starts to empty, with the most strongly correlated states (those experiencing the most repulsion or flattest dispersion) emptying first.

A further increase in the Peierls spacing results in the band structures and densities of states of Figure 8, which adopt the now familiar form of a ferromagnetic structure similar to the $\frac{3}{4}$ and EXP structures. In this form, the (erstwhile) bonding band stays relatively unchanged; its dispersion is almost identical to that of Figures 5-7. However, the spin down component of this band has now almost completely emptied, with the remaining states, which correspond to almost two electrons, filling the lowest lying conduction bands.

A clearer picture can be obtained by converting the peaks of the density of states of Figures 2c and 3b into charge densities, as Figure 9 illustrates. Figures 9a and 9b exhibit slices of the  charge density of the a) $d$-band peak of the $\frac{1}{2}$ structure (highlighted in blue in Figure 3b) and the b) $d$-band (lower energy) peak of the EXP structure (blue highlight of Figure 2c) in the $(0\bar{1}1)$ plane. The EXP structure is used here for comparison as the up and down spin states do not overlap in the conduction band, allowing their charge densities to be resolved. The EXP band structure and its comparison to that of Figure 8 is presented in the Appendix. The charge densities are clearly very similar, consisting of high density delocalized between the Peierls paired vanadium atoms, with high density between the pairs, providing further confirmation of the Peierls bonding scenario. The magnitude of the interstitial density of the EXP structure is slightly lower, due to the fact that this band corresponds to half as many electrons than in the $\frac{1}{2}$ structure, and the overlapping nuclear potential is lower in magnitude. However, in Figure 9c, the high energy $d$-band peak of the EXP structure (red highlight of Figure 2c) exhibits rather different character. The charge density is suggestive of an atomic-like orbital, highly localized on the vanadium atoms. Therefore, in the EXP structure each Peierls pair contains one spin up electron inhabiting a molecular (bonding) orbital delocalized across the pair, and one electron with a wavefunction concentrated in atomic-like orbitals centered on the vanadium atoms.  

Such splitting of the $d$-band states from a single, spin-paired molecular orbital into two separate spin up orbitals is clearly a manifestation of Hund's first rule. The increase in the distance between the Peierls paired vanadium atoms decreases the nuclear potential overlap, thereby decreasing the stabilization of the bonding electrons with respect to their mutual repulsion. As a consequence, this repulsion forces the splitting of the occupation into two separate, orthogonal, orbitals with spin alignment blocking hopping between the bonding and non-bonding orbitals due to the Pauli principle, decreasing repulsion. 

A comparison of the low energy excitations of the $\frac{3}{4}$ and EXP structures (Figures 3c and 2c respectively) indicates that as the inter-vanadium distance of the Peierls pairs continues to increase, the spin down excitations shift to higher energy, while the spin up peak, which forms a doublet with the occupied spin up states of Figure 9c, remains unchanged. Therefore, as the nuclear potential overlap between the Peierls paired vanadium atoms is increased further, it is less able to stabilize the repulsion of spin paired bonding electrons filling the states of Figure 9b, and it shifts to higher energy. Adding a spin up low energy excitation into the $\frac{3}{4}$ or the EXP structure results in three spin up electrons spread over the four atomic-like orbitals of Figure 9c per unit cell. Thus, hopping will be slightly inhibited, however there still exists a hole for electrons to move into, lowering their kinetic energies. Figure 9d illustrates the charge density of the green shaded peak of Figure 2c, and while the lobes of the orbitals differ slightly, the nodes indicate that the wavefunction is identical to the filled band of the EXP structure. 

Of interest is the fact that the splitting of the EXP structure at the Fermi level in the GW data corresponds to identical charge densities being split into a doublet approximately symmetric about the Fermi level, upon the inclusion of non-local correlations in a quasiparticle description. Therefore the added spin up electron sees a shift to higher energy once the interactions with the valence electrons are properly taken into account. This splitting suggests that these two structures may exhibit Mott-Hubbard character, with on-site repulsion significantly influencing quasiparticle properties. However, single shot G$_0$W$_0$ calculations such as these are not able to properly account for such interactions, as they are based on single particle equations and represent only a first order correction, and therefore the magnitude of this splitting may be significantly understated.

\section{Conclusion}
Spin-resolved GW calculations reveal that the magnetic structure of M1 vanadium dioxide depends most significantly on the inter-vanadium spacing of the Peierls pairs. Perturbations of other structural parameters manifested almost no effects on spin ordering. Starting from an antiferromagnetically ordered structure with a band gap in agreement with experiment (0.65 eV) and increasing the Peierls pairing distance results in a gradual transition to ferromagnetic ordering. The band gap and spin structure is revealed to be stable to slight perturbations, however further increasing the Peierls spacing results in the overlapping nuclear potential of the pairs weakening, such that an electron is released into the conduction band. In the ground state this electron is spin aligned with the electron occupying the Peierls band, as this occupation of orthogonal wavefunctions and spin alignment results in the minimization of electron repulsion: a manifestation of Hund's first rule, creating a ferromagnetic structure.

\section{Appendix}
Figure 10 presents the band structure of the EXP structure as a comparison to that of Figure 8. Inspection of the band dispersions of the up spin states in the GW data (filled circles) of both Figures suggests that they are virtually identical. However, in the EXP structure the spin down states are shifted to higher energy, such that the leading edge of the spin down conduction band does not fall in the same energy range as the peak corresponding to the conduction band of the spin up states. Therefore, the spin up states appear to change very little passing from Figure 8 to the EXP structure. The shift in the down states however, allows the up spin conduction band of the EXP structure to be converted to charge density without inclusion of any spin down states. For this reason, Figure 9 of the main text uses charge densities of the EXP structure.
\begin{figure}[h!]
\centering
\includegraphics[width=\columnwidth]{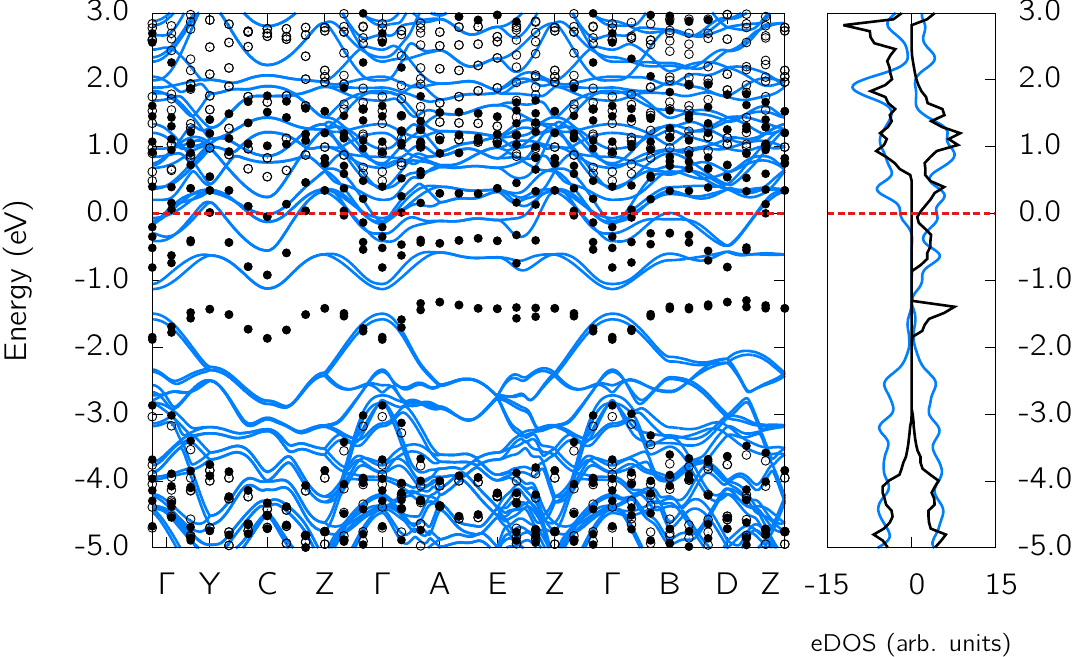}
\caption{Electronic bands of the EXP structure. Symbol and line conventions are as per Figure 5.}
\end{figure}

\section{Acknowledgements}
JMB, DWD, PSC, JSS and SPR acknowledge support from the National Computational Institute.

\bibliography{F:/GWApproximation/Bibliography/library}

\begin{thebibliography}{33}%
\makeatletter
\providecommand \@ifxundefined [1]{%
 \@ifx{#1\undefined}
}%
\providecommand \@ifnum [1]{%
 \ifnum #1\expandafter \@firstoftwo
 \else \expandafter \@secondoftwo
 \fi
}%
\providecommand \@ifx [1]{%
 \ifx #1\expandafter \@firstoftwo
 \else \expandafter \@secondoftwo
 \fi
}%
\providecommand \natexlab [1]{#1}%
\providecommand \enquote  [1]{``#1''}%
\providecommand \bibnamefont  [1]{#1}%
\providecommand \bibfnamefont [1]{#1}%
\providecommand \citenamefont [1]{#1}%
\providecommand \href@noop [0]{\@secondoftwo}%
\providecommand \href [0]{\begingroup \@sanitize@url \@href}%
\providecommand \@href[1]{\@@startlink{#1}\@@href}%
\providecommand \@@href[1]{\endgroup#1\@@endlink}%
\providecommand \@sanitize@url [0]{\catcode `\\12\catcode `\$12\catcode
  `\&12\catcode `\#12\catcode `\^12\catcode `\_12\catcode `\%12\relax}%
\providecommand \@@startlink[1]{}%
\providecommand \@@endlink[0]{}%
\providecommand \url  [0]{\begingroup\@sanitize@url \@url }%
\providecommand \@url [1]{\endgroup\@href {#1}{\urlprefix }}%
\providecommand \urlprefix  [0]{URL }%
\providecommand \Eprint [0]{\href }%
\providecommand \doibase [0]{http://dx.doi.org/}%
\providecommand \selectlanguage [0]{\@gobble}%
\providecommand \bibinfo  [0]{\@secondoftwo}%
\providecommand \bibfield  [0]{\@secondoftwo}%
\providecommand \translation [1]{[#1]}%
\providecommand \BibitemOpen [0]{}%
\providecommand \bibitemStop [0]{}%
\providecommand \bibitemNoStop [0]{.\EOS\space}%
\providecommand \EOS [0]{\spacefactor3000\relax}%
\providecommand \BibitemShut  [1]{\csname bibitem#1\endcsname}%
\let\auto@bib@innerbib\@empty
\bibitem [{\citenamefont {Dagotto}(2005)}]{Dagotto2005}%
  \BibitemOpen
  \bibfield  {author} {\bibinfo {author} {\bibfnamefont {E.}~\bibnamefont
  {Dagotto}},\ }\href {\doibase 10.1126/science.1107559} {\bibfield  {journal}
  {\bibinfo  {journal} {Science}\ }\textbf {\bibinfo {volume} {309}},\ \bibinfo
  {pages} {257} (\bibinfo {year} {2005})}\BibitemShut {NoStop}%
\bibitem [{\citenamefont {Basov}\ \emph {et~al.}(2011)\citenamefont {Basov},
  \citenamefont {Averitt}, \citenamefont {van~der Marel}, \citenamefont
  {Dressel},\ and\ \citenamefont {Haule}}]{Basov2011}%
  \BibitemOpen
  \bibfield  {author} {\bibinfo {author} {\bibfnamefont {D.~N.}\ \bibnamefont
  {Basov}}, \bibinfo {author} {\bibfnamefont {R.~D.}\ \bibnamefont {Averitt}},
  \bibinfo {author} {\bibfnamefont {D.}~\bibnamefont {van~der Marel}}, \bibinfo
  {author} {\bibfnamefont {M.}~\bibnamefont {Dressel}}, \ and\ \bibinfo
  {author} {\bibfnamefont {K.}~\bibnamefont {Haule}},\ }\href {\doibase
  10.1103/RevModPhys.83.471} {\bibfield  {journal} {\bibinfo  {journal} {Rev.
  Mod. Phys.}\ }\textbf {\bibinfo {volume} {83}},\ \bibinfo {pages} {471}
  (\bibinfo {year} {2011})}\BibitemShut {NoStop}%
\bibitem [{\citenamefont {Morin}(1959)}]{Morin1959}%
  \BibitemOpen
  \bibfield  {author} {\bibinfo {author} {\bibfnamefont {F.~J.}\ \bibnamefont
  {Morin}},\ }\href@noop {} {\bibfield  {journal} {\bibinfo  {journal} {Phys.
  Rev. Lett.}\ }\textbf {\bibinfo {volume} {3}},\ \bibinfo {pages} {2}
  (\bibinfo {year} {1959})}\BibitemShut {NoStop}%
\bibitem [{\citenamefont {Goodenough}(1971)}]{Goodenough1971}%
  \BibitemOpen
  \bibfield  {author} {\bibinfo {author} {\bibfnamefont {J.~B.}\ \bibnamefont
  {Goodenough}},\ }\href@noop {} {\bibfield  {journal} {\bibinfo  {journal} {J.
  Solid State Chem.}\ }\textbf {\bibinfo {volume} {3}},\ \bibinfo {pages} {490}
  (\bibinfo {year} {1971})}\BibitemShut {NoStop}%
\bibitem [{\citenamefont {Mott}\ and\ \citenamefont
  {Friedman}(1974)}]{Mott1974}%
  \BibitemOpen
  \bibfield  {author} {\bibinfo {author} {\bibfnamefont {N.~F.}\ \bibnamefont
  {Mott}}\ and\ \bibinfo {author} {\bibfnamefont {L.}~\bibnamefont
  {Friedman}},\ }\href@noop {} {\bibfield  {journal} {\bibinfo  {journal}
  {Philos. Mag.}\ }\textbf {\bibinfo {volume} {30}},\ \bibinfo {pages} {389}
  (\bibinfo {year} {1974})}\BibitemShut {NoStop}%
\bibitem [{\citenamefont {Pouget}\ \emph {et~al.}(1975)\citenamefont {Pouget},
  \citenamefont {Launois}, \citenamefont {D'Haenens}, \citenamefont {Merenda},\
  and\ \citenamefont {{Rice, Tim}}}]{Pouget1975}%
  \BibitemOpen
  \bibfield  {author} {\bibinfo {author} {\bibfnamefont {J.~P.}\ \bibnamefont
  {Pouget}}, \bibinfo {author} {\bibfnamefont {H.}~\bibnamefont {Launois}},
  \bibinfo {author} {\bibfnamefont {J.~P.}\ \bibnamefont {D'Haenens}}, \bibinfo
  {author} {\bibfnamefont {P.}~\bibnamefont {Merenda}}, \ and\ \bibinfo
  {author} {\bibfnamefont {M.}~\bibnamefont {{Rice, Tim}}},\ }\href@noop {}
  {\bibfield  {journal} {\bibinfo  {journal} {Phys. Rev. Lett.}\ }\textbf
  {\bibinfo {volume} {35}},\ \bibinfo {pages} {873} (\bibinfo {year}
  {1975})}\BibitemShut {NoStop}%
\bibitem [{\citenamefont {Zylberstein}\ and\ \citenamefont
  {Mott}(1975)}]{Zylberstein1975}%
  \BibitemOpen
  \bibfield  {author} {\bibinfo {author} {\bibfnamefont {A.}~\bibnamefont
  {Zylberstein}}\ and\ \bibinfo {author} {\bibfnamefont {N.~F.}\ \bibnamefont
  {Mott}},\ }\href@noop {} {\bibfield  {journal} {\bibinfo  {journal} {Phys.
  Rev. B}\ }\textbf {\bibinfo {volume} {11}},\ \bibinfo {pages} {4383}
  (\bibinfo {year} {1975})}\BibitemShut {NoStop}%
\bibitem [{\citenamefont {Imada}\ \emph {et~al.}(1998)\citenamefont {Imada},
  \citenamefont {Fujimori},\ and\ \citenamefont {Tokura}}]{Imada1998}%
  \BibitemOpen
  \bibfield  {author} {\bibinfo {author} {\bibfnamefont {M.}~\bibnamefont
  {Imada}}, \bibinfo {author} {\bibfnamefont {A.}~\bibnamefont {Fujimori}}, \
  and\ \bibinfo {author} {\bibfnamefont {Y.}~\bibnamefont {Tokura}},\ }\href
  {\doibase 10.1103/RevModPhys.70.1039} {\bibfield  {journal} {\bibinfo
  {journal} {Rev. Mod. Phys.}\ }\textbf {\bibinfo {volume} {70}},\ \bibinfo
  {pages} {1039} (\bibinfo {year} {1998})}\BibitemShut {NoStop}%
\bibitem [{\citenamefont {Kotliar}\ \emph {et~al.}(2006)\citenamefont
  {Kotliar}, \citenamefont {Savrasov}, \citenamefont {Haule}, \citenamefont
  {Oudovenko}, \citenamefont {Parcollet},\ and\ \citenamefont
  {Marianetti}}]{Kotliar2006}%
  \BibitemOpen
  \bibfield  {author} {\bibinfo {author} {\bibfnamefont {G.}~\bibnamefont
  {Kotliar}}, \bibinfo {author} {\bibfnamefont {S.}~\bibnamefont {Savrasov}},
  \bibinfo {author} {\bibfnamefont {K.}~\bibnamefont {Haule}}, \bibinfo
  {author} {\bibfnamefont {V.}~\bibnamefont {Oudovenko}}, \bibinfo {author}
  {\bibfnamefont {O.}~\bibnamefont {Parcollet}}, \ and\ \bibinfo {author}
  {\bibfnamefont {C.}~\bibnamefont {Marianetti}},\ }\href {\doibase
  10.1103/RevModPhys.78.865} {\bibfield  {journal} {\bibinfo  {journal} {Rev.
  Mod. Phys.}\ }\textbf {\bibinfo {volume} {78}},\ \bibinfo {pages} {865}
  (\bibinfo {year} {2006})}\BibitemShut {NoStop}%
\bibitem [{\citenamefont {Hedin}(1965)}]{Hedin1965}%
  \BibitemOpen
  \bibfield  {author} {\bibinfo {author} {\bibfnamefont {L.}~\bibnamefont
  {Hedin}},\ }\href@noop {} {\bibfield  {journal} {\bibinfo  {journal} {Phys.
  Rev.}\ }\textbf {\bibinfo {volume} {139}},\ \bibinfo {pages} {796} (\bibinfo
  {year} {1965})}\BibitemShut {NoStop}%
\bibitem [{\citenamefont {Tomczak}\ \emph {et~al.}(2008)\citenamefont
  {Tomczak}, \citenamefont {Aryasetiawan},\ and\ \citenamefont
  {Biermann}}]{Tomczak2008}%
  \BibitemOpen
  \bibfield  {author} {\bibinfo {author} {\bibfnamefont {J.~M.}\ \bibnamefont
  {Tomczak}}, \bibinfo {author} {\bibfnamefont {F.}~\bibnamefont
  {Aryasetiawan}}, \ and\ \bibinfo {author} {\bibfnamefont {S.}~\bibnamefont
  {Biermann}},\ }\href {\doibase 10.1103/PhysRevB.78.115103} {\bibfield
  {journal} {\bibinfo  {journal} {Phys. Rev. B}\ }\textbf {\bibinfo {volume}
  {78}},\ \bibinfo {pages} {115103} (\bibinfo {year} {2008})}\BibitemShut
  {NoStop}%
\bibitem [{\citenamefont {Gatti}\ \emph {et~al.}(2007)\citenamefont {Gatti},
  \citenamefont {Bruneval}, \citenamefont {Olevano},\ and\ \citenamefont
  {Reining}}]{Gatti2007}%
  \BibitemOpen
  \bibfield  {author} {\bibinfo {author} {\bibfnamefont {M.}~\bibnamefont
  {Gatti}}, \bibinfo {author} {\bibfnamefont {F.}~\bibnamefont {Bruneval}},
  \bibinfo {author} {\bibfnamefont {V.}~\bibnamefont {Olevano}}, \ and\
  \bibinfo {author} {\bibfnamefont {L.}~\bibnamefont {Reining}},\ }\href
  {\doibase 10.1103/PhysRevLett.99.266402} {\bibfield  {journal} {\bibinfo
  {journal} {Phys. Rev. Lett.}\ }\textbf {\bibinfo {volume} {99}},\ \bibinfo
  {pages} {266402} (\bibinfo {year} {2007})}\BibitemShut {NoStop}%
\bibitem [{\citenamefont {Belozerov}\ \emph {et~al.}(2012)\citenamefont
  {Belozerov}, \citenamefont {Korotin}, \citenamefont {Anisimov},\ and\
  \citenamefont {Poteryaev}}]{Belozerov2012}%
  \BibitemOpen
  \bibfield  {author} {\bibinfo {author} {\bibfnamefont {A.~S.}\ \bibnamefont
  {Belozerov}}, \bibinfo {author} {\bibfnamefont {M.~A.}\ \bibnamefont
  {Korotin}}, \bibinfo {author} {\bibfnamefont {V.~I.}\ \bibnamefont
  {Anisimov}}, \ and\ \bibinfo {author} {\bibfnamefont {A.~I.}\ \bibnamefont
  {Poteryaev}},\ }\href {\doibase 10.1103/PhysRevB.85.045109} {\bibfield
  {journal} {\bibinfo  {journal} {Phys. Rev. B}\ }\textbf {\bibinfo {volume}
  {85}},\ \bibinfo {pages} {045109} (\bibinfo {year} {2012})}\BibitemShut
  {NoStop}%
\bibitem [{\citenamefont {Peierls}(1955)}]{Peierls1955}%
  \BibitemOpen
  \bibfield  {author} {\bibinfo {author} {\bibfnamefont {R.~E.}\ \bibnamefont
  {Peierls}},\ }\href@noop {} {\emph {\bibinfo {title} {{Quantum Theory of
  Solids}}}}\ (\bibinfo  {publisher} {Oxford University Press},\ \bibinfo
  {address} {Oxford},\ \bibinfo {year} {1955})\ p.\ \bibinfo {pages}
  {108}\BibitemShut {NoStop}%
\bibitem [{\citenamefont {Wu}\ \emph {et~al.}(2006)\citenamefont {Wu},
  \citenamefont {Gu}, \citenamefont {Guiton}, \citenamefont {Leon},
  \citenamefont {Ouyang},\ and\ \citenamefont {Park}}]{Wu2006}%
  \BibitemOpen
  \bibfield  {author} {\bibinfo {author} {\bibfnamefont {J.}~\bibnamefont
  {Wu}}, \bibinfo {author} {\bibfnamefont {Q.}~\bibnamefont {Gu}}, \bibinfo
  {author} {\bibfnamefont {B.~S.}\ \bibnamefont {Guiton}}, \bibinfo {author}
  {\bibfnamefont {N.~P.~D.}\ \bibnamefont {Leon}}, \bibinfo {author}
  {\bibfnamefont {L.}~\bibnamefont {Ouyang}}, \ and\ \bibinfo {author}
  {\bibfnamefont {H.}~\bibnamefont {Park}},\ }\href@noop {} {\bibfield
  {journal} {\bibinfo  {journal} {Nano Lett.}\ }\textbf {\bibinfo {volume}
  {6}},\ \bibinfo {pages} {2313} (\bibinfo {year} {2006})}\BibitemShut
  {NoStop}%
\bibitem [{\citenamefont {Zhou}\ \emph {et~al.}(2008)\citenamefont {Zhou},
  \citenamefont {Gu}, \citenamefont {Fei}, \citenamefont {Mai}, \citenamefont
  {Gao}, \citenamefont {Yang}, \citenamefont {Bao},\ and\ \citenamefont
  {Wang}}]{Zhou2008}%
  \BibitemOpen
  \bibfield  {author} {\bibinfo {author} {\bibfnamefont {J.}~\bibnamefont
  {Zhou}}, \bibinfo {author} {\bibfnamefont {Y.}~\bibnamefont {Gu}}, \bibinfo
  {author} {\bibfnamefont {P.}~\bibnamefont {Fei}}, \bibinfo {author}
  {\bibfnamefont {W.}~\bibnamefont {Mai}}, \bibinfo {author} {\bibfnamefont
  {Y.}~\bibnamefont {Gao}}, \bibinfo {author} {\bibfnamefont {R.}~\bibnamefont
  {Yang}}, \bibinfo {author} {\bibfnamefont {G.}~\bibnamefont {Bao}}, \ and\
  \bibinfo {author} {\bibfnamefont {Z.~L.}\ \bibnamefont {Wang}},\ }\href
  {\doibase 10.1021/nl802367t} {\bibfield  {journal} {\bibinfo  {journal} {Nano
  Lett.}\ }\textbf {\bibinfo {volume} {8}},\ \bibinfo {pages} {3035} (\bibinfo
  {year} {2008})}\BibitemShut {NoStop}%
\bibitem [{\citenamefont {Sohn}\ \emph {et~al.}(2009)\citenamefont {Sohn},
  \citenamefont {Joo}, \citenamefont {Ahn}, \citenamefont {Lee}, \citenamefont
  {Porter}, \citenamefont {Kim}, \citenamefont {Kang},\ and\ \citenamefont
  {Welland}}]{Sohn2009}%
  \BibitemOpen
  \bibfield  {author} {\bibinfo {author} {\bibfnamefont {J.~I.}\ \bibnamefont
  {Sohn}}, \bibinfo {author} {\bibfnamefont {H.~J.}\ \bibnamefont {Joo}},
  \bibinfo {author} {\bibfnamefont {D.}~\bibnamefont {Ahn}}, \bibinfo {author}
  {\bibfnamefont {H.~H.}\ \bibnamefont {Lee}}, \bibinfo {author} {\bibfnamefont
  {A.~E.}\ \bibnamefont {Porter}}, \bibinfo {author} {\bibfnamefont
  {K.}~\bibnamefont {Kim}}, \bibinfo {author} {\bibfnamefont {D.~J.}\
  \bibnamefont {Kang}}, \ and\ \bibinfo {author} {\bibfnamefont {M.~E.}\
  \bibnamefont {Welland}},\ }\href {\doibase 10.1021/nl900841k} {\bibfield
  {journal} {\bibinfo  {journal} {Nano Lett.}\ }\textbf {\bibinfo {volume}
  {9}},\ \bibinfo {pages} {3392} (\bibinfo {year} {2009})}\BibitemShut
  {NoStop}%
\bibitem [{\citenamefont {Guo}\ \emph {et~al.}(2011)\citenamefont {Guo},
  \citenamefont {Chen}, \citenamefont {Oh}, \citenamefont {Wang}, \citenamefont
  {Dejoie}, \citenamefont {{Syed Asif}}, \citenamefont {Warren}, \citenamefont
  {Shan}, \citenamefont {Wu},\ and\ \citenamefont {Minor}}]{Guo2011}%
  \BibitemOpen
  \bibfield  {author} {\bibinfo {author} {\bibfnamefont {H.}~\bibnamefont
  {Guo}}, \bibinfo {author} {\bibfnamefont {K.}~\bibnamefont {Chen}}, \bibinfo
  {author} {\bibfnamefont {Y.}~\bibnamefont {Oh}}, \bibinfo {author}
  {\bibfnamefont {K.}~\bibnamefont {Wang}}, \bibinfo {author} {\bibfnamefont
  {C.}~\bibnamefont {Dejoie}}, \bibinfo {author} {\bibfnamefont
  {S.}~\bibnamefont {{Syed Asif}}}, \bibinfo {author} {\bibfnamefont {O.~L.}\
  \bibnamefont {Warren}}, \bibinfo {author} {\bibfnamefont {Z.~W.}\
  \bibnamefont {Shan}}, \bibinfo {author} {\bibfnamefont {J.}~\bibnamefont
  {Wu}}, \ and\ \bibinfo {author} {\bibfnamefont {M.}~\bibnamefont {Minor}},\
  }\href {\doibase 10.1021/nl201460v} {\bibfield  {journal} {\bibinfo
  {journal} {Nano Lett.}\ }\textbf {\bibinfo {volume} {11}},\ \bibinfo {pages}
  {3207} (\bibinfo {year} {2011})}\BibitemShut {NoStop}%
\bibitem [{\citenamefont {Nakano}\ \emph {et~al.}(2012)\citenamefont {Nakano},
  \citenamefont {Shibuya}, \citenamefont {Okuyama}, \citenamefont {Hatano},
  \citenamefont {Ono}, \citenamefont {Kawasaki}, \citenamefont {Iwasa},\ and\
  \citenamefont {Tokura}}]{Nakano2012}%
  \BibitemOpen
  \bibfield  {author} {\bibinfo {author} {\bibfnamefont {M.}~\bibnamefont
  {Nakano}}, \bibinfo {author} {\bibfnamefont {K.}~\bibnamefont {Shibuya}},
  \bibinfo {author} {\bibfnamefont {D.}~\bibnamefont {Okuyama}}, \bibinfo
  {author} {\bibfnamefont {T.}~\bibnamefont {Hatano}}, \bibinfo {author}
  {\bibfnamefont {S.}~\bibnamefont {Ono}}, \bibinfo {author} {\bibfnamefont
  {M.}~\bibnamefont {Kawasaki}}, \bibinfo {author} {\bibfnamefont
  {Y.}~\bibnamefont {Iwasa}}, \ and\ \bibinfo {author} {\bibfnamefont
  {Y.}~\bibnamefont {Tokura}},\ }\href {\doibase 10.1038/nature11296}
  {\bibfield  {journal} {\bibinfo  {journal} {Nature}\ }\textbf {\bibinfo
  {volume} {487}},\ \bibinfo {pages} {459} (\bibinfo {year}
  {2012})}\BibitemShut {NoStop}%
\bibitem [{\citenamefont {Andersson}(1954)}]{Andersson1954}%
  \BibitemOpen
  \bibfield  {author} {\bibinfo {author} {\bibfnamefont {G.}~\bibnamefont
  {Andersson}},\ }\href@noop {} {\bibfield  {journal} {\bibinfo  {journal}
  {Acta Chem. Scand.}\ }\textbf {\bibinfo {volume} {8}},\ \bibinfo {pages}
  {1599} (\bibinfo {year} {1954})}\BibitemShut {NoStop}%
\bibitem [{\citenamefont {Kresse}\ and\ \citenamefont
  {Furthm\"{u}ller}(1996)}]{Kresse1996}%
  \BibitemOpen
  \bibfield  {author} {\bibinfo {author} {\bibfnamefont {G.}~\bibnamefont
  {Kresse}}\ and\ \bibinfo {author} {\bibfnamefont {J.}~\bibnamefont
  {Furthm\"{u}ller}},\ }\href {http://www.ncbi.nlm.nih.gov/pubmed/9984901}
  {\bibfield  {journal} {\bibinfo  {journal} {Phys. Rev. B}\ }\textbf {\bibinfo
  {volume} {54}},\ \bibinfo {pages} {11169} (\bibinfo {year}
  {1996})}\BibitemShut {NoStop}%
\bibitem [{\citenamefont {Perdew}\ \emph {et~al.}(1996)\citenamefont {Perdew},
  \citenamefont {Burke},\ and\ \citenamefont {Ernzerhof}}]{Perdew1996}%
  \BibitemOpen
  \bibfield  {author} {\bibinfo {author} {\bibfnamefont {J.~P.}\ \bibnamefont
  {Perdew}}, \bibinfo {author} {\bibfnamefont {K.}~\bibnamefont {Burke}}, \
  and\ \bibinfo {author} {\bibfnamefont {M.}~\bibnamefont {Ernzerhof}},\ }\href
  {http://www.ncbi.nlm.nih.gov/pubmed/10062328} {\bibfield  {journal} {\bibinfo
   {journal} {Phys. Rev. Lett.}\ }\textbf {\bibinfo {volume} {77}},\ \bibinfo
  {pages} {3865} (\bibinfo {year} {1996})}\BibitemShut {NoStop}%
\bibitem [{\citenamefont {Methfessel}\ and\ \citenamefont
  {Paxton}(1989)}]{Methfessel1989}%
  \BibitemOpen
  \bibfield  {author} {\bibinfo {author} {\bibfnamefont {M.}~\bibnamefont
  {Methfessel}}\ and\ \bibinfo {author} {\bibfnamefont {A.~T.}\ \bibnamefont
  {Paxton}},\ }\href@noop {} {\bibfield  {journal} {\bibinfo  {journal} {Phys.
  Rev. B}\ }\textbf {\bibinfo {volume} {40}},\ \bibinfo {pages} {3616}
  (\bibinfo {year} {1989})}\BibitemShut {NoStop}%
\bibitem [{\citenamefont {Bloechl}\ \emph {et~al.}(1994)\citenamefont
  {Bloechl}, \citenamefont {Jepsen},\ and\ \citenamefont
  {Andersen}}]{Bloechl1994}%
  \BibitemOpen
  \bibfield  {author} {\bibinfo {author} {\bibfnamefont {P.~E.}\ \bibnamefont
  {Bloechl}}, \bibinfo {author} {\bibfnamefont {O.}~\bibnamefont {Jepsen}}, \
  and\ \bibinfo {author} {\bibfnamefont {O.~K.}\ \bibnamefont {Andersen}},\
  }\href@noop {} {\bibfield  {journal} {\bibinfo  {journal} {Phys. Rev. B}\
  }\textbf {\bibinfo {volume} {49}},\ \bibinfo {pages} {16223} (\bibinfo {year}
  {1994})}\BibitemShut {NoStop}%
\bibitem [{\citenamefont {Shishkin}\ and\ \citenamefont
  {Kresse}(2006)}]{Shishkin2006}%
  \BibitemOpen
  \bibfield  {author} {\bibinfo {author} {\bibfnamefont {M.}~\bibnamefont
  {Shishkin}}\ and\ \bibinfo {author} {\bibfnamefont {G.}~\bibnamefont
  {Kresse}},\ }\href {\doibase 10.1103/PhysRevB.74.035101} {\bibfield
  {journal} {\bibinfo  {journal} {Phys. Rev. B}\ }\textbf {\bibinfo {volume}
  {74}},\ \bibinfo {pages} {035101} (\bibinfo {year} {2006})}\BibitemShut
  {NoStop}%
\bibitem [{\citenamefont {Kohn}\ and\ \citenamefont {Sham}(1965)}]{Kohn1965}%
  \BibitemOpen
  \bibfield  {author} {\bibinfo {author} {\bibfnamefont {W.}~\bibnamefont
  {Kohn}}\ and\ \bibinfo {author} {\bibfnamefont {L.~J.}\ \bibnamefont
  {Sham}},\ }\href@noop {} {\bibfield  {journal} {\bibinfo  {journal} {Phys.
  Rev.}\ }\textbf {\bibinfo {volume} {140}},\ \bibinfo {pages} {1133} (\bibinfo
  {year} {1965})}\BibitemShut {NoStop}%
\bibitem [{\citenamefont {Monkhorst}\ and\ \citenamefont
  {Pack}(1976)}]{Monkhorst1976}%
  \BibitemOpen
  \bibfield  {author} {\bibinfo {author} {\bibfnamefont {H.~J.}\ \bibnamefont
  {Monkhorst}}\ and\ \bibinfo {author} {\bibfnamefont {J.~D.}\ \bibnamefont
  {Pack}},\ }\href@noop {} {\bibfield  {journal} {\bibinfo  {journal} {Phys.
  Rev. B}\ }\textbf {\bibinfo {volume} {13}},\ \bibinfo {pages} {5188}
  (\bibinfo {year} {1976})}\BibitemShut {NoStop}%
\bibitem [{\citenamefont {Guiliani}\ and\ \citenamefont
  {Vignale}(2005)}]{Guiliani2005}%
  \BibitemOpen
  \bibfield  {author} {\bibinfo {author} {\bibfnamefont {G.~F.}\ \bibnamefont
  {Guiliani}}\ and\ \bibinfo {author} {\bibfnamefont {G.}~\bibnamefont
  {Vignale}},\ }\href@noop {} {\emph {\bibinfo {title} {{Quantum Theory of the
  Electron Liquid}}}}\ (\bibinfo  {publisher} {Cambridge University Press},\
  \bibinfo {address} {New York},\ \bibinfo {year} {2005})\ p.\ \bibinfo {pages}
  {296}\BibitemShut {NoStop}%
\bibitem [{\citenamefont {Hybertsen}\ and\ \citenamefont
  {Louie}(1987)}]{Hybertsen1987}%
  \BibitemOpen
  \bibfield  {author} {\bibinfo {author} {\bibfnamefont {M.~S.}\ \bibnamefont
  {Hybertsen}}\ and\ \bibinfo {author} {\bibfnamefont {S.~G.}\ \bibnamefont
  {Louie}},\ }\href@noop {} {\bibfield  {journal} {\bibinfo  {journal} {Phys.
  Rev. B}\ }\textbf {\bibinfo {volume} {35}},\ \bibinfo {pages} {5585}
  (\bibinfo {year} {1987})}\BibitemShut {NoStop}%
\bibitem [{\citenamefont {Wentzcovitch}\ \emph {et~al.}(1994)\citenamefont
  {Wentzcovitch}, \citenamefont {Schulz},\ and\ \citenamefont
  {Allen}}]{Wentzcovich1994}%
  \BibitemOpen
  \bibfield  {author} {\bibinfo {author} {\bibfnamefont {R.~M.}\ \bibnamefont
  {Wentzcovitch}}, \bibinfo {author} {\bibfnamefont {W.~W.}\ \bibnamefont
  {Schulz}}, \ and\ \bibinfo {author} {\bibfnamefont {P.~B.}\ \bibnamefont
  {Allen}},\ }\href@noop {} {\bibfield  {journal} {\bibinfo  {journal} {Phys.
  Rev. Lett.}\ }\textbf {\bibinfo {volume} {72}},\ \bibinfo {pages} {3389}
  (\bibinfo {year} {1994})}\BibitemShut {NoStop}%
\bibitem [{\citenamefont {Eyert}(2002)}]{Eyert2002}%
  \BibitemOpen
  \bibfield  {author} {\bibinfo {author} {\bibfnamefont {V.}~\bibnamefont
  {Eyert}},\ }\href {http://arxiv.org/abs/cond-mat/0210558} {\bibfield
  {journal} {\bibinfo  {journal} {Ann. Phys.}\ }\textbf {\bibinfo {volume}
  {11}},\ \bibinfo {pages} {650} (\bibinfo {year} {2002})}\BibitemShut
  {NoStop}%
\bibitem [{\citenamefont {Muraoka}\ \emph {et~al.}(2002)\citenamefont
  {Muraoka}, \citenamefont {Ueda},\ and\ \citenamefont {Hiroi}}]{Muraoka2002}%
  \BibitemOpen
  \bibfield  {author} {\bibinfo {author} {\bibfnamefont {Y.}~\bibnamefont
  {Muraoka}}, \bibinfo {author} {\bibfnamefont {Y.}~\bibnamefont {Ueda}}, \
  and\ \bibinfo {author} {\bibfnamefont {Z.}~\bibnamefont {Hiroi}},\
  }\href@noop {} {\bibfield  {journal} {\bibinfo  {journal} {J. Phys. Chem.
  Solids}\ }\textbf {\bibinfo {volume} {63}},\ \bibinfo {pages} {965} (\bibinfo
  {year} {2002})}\BibitemShut {NoStop}%
\bibitem [{\citenamefont {Gregg}\ and\ \citenamefont
  {Bowman}(1997)}]{Gregg1997}%
  \BibitemOpen
  \bibfield  {author} {\bibinfo {author} {\bibfnamefont {J.~M.}\ \bibnamefont
  {Gregg}}\ and\ \bibinfo {author} {\bibfnamefont {R.~M.}\ \bibnamefont
  {Bowman}},\ }\href {\doibase 10.1063/1.120469} {\bibfield  {journal}
  {\bibinfo  {journal} {Appl. Phys. Lett.}\ }\textbf {\bibinfo {volume} {71}},\
  \bibinfo {pages} {3649} (\bibinfo {year} {1997})}\BibitemShut {NoStop}%
\end{thebibliography}%
\end{document}